\documentclass[journal]{IEEEtran}
\usepackage{xcolor,soul,framed} 
\usepackage{graphicx}
\usepackage{float}
\usepackage{subfigure}
\usepackage{graphicx}
\usepackage[flushleft]{threeparttable}
\usepackage{adjustbox}
\usepackage{color}

\colorlet{shadecolor}{yellow}

\usepackage{array}
\usepackage{mdwmath}
\usepackage{mdwtab}
\usepackage{eqparbox}
\usepackage{url}

\begin{document}
    \title{Interacting with New York City Data by HoloLens through Remote Rendering}
\author{Zijian Long, Haiwei Dong,~\IEEEmembership{Senior Member,~IEEE}, and Abdulmotaleb El Saddik,~\IEEEmembership{Fellow,~IEEE}}


\maketitle

\begin{abstract}
In the digital era, Extended Reality (XR) is considered the next frontier. However, XR systems are computationally intensive, and they must be implemented within strict latency constraints. Thus, XR devices with finite computing resources are limited in terms of quality of experience (QoE) they can offer, particularly in cases of big 3D data. This problem can be effectively addressed by offloading the highly intensive rendering tasks to a remote server. Therefore, we proposed a remote rendering enabled XR system that presents the 3D city model of New York City on the Microsoft HoloLens. Experimental results indicate that remote rendering outperforms local rendering for the New York City model with significant improvement in average QoE by at least 21\%. Additionally, we clarified the network traffic pattern in the proposed XR system developed under the OpenXR standard.

\end{abstract}


%
\IEEEpeerreviewmaketitle


\section{HoloLens and Remote Rendering}
Extended reality is a term referring to all real-and-virtual combined environments and human-machine interactions \cite{fast2018testing}. It subsumes the entire spectrum of realities assisted by immersive technology such as Virtual Reality (VR), Augmented Reality (AR) and Mixed Reality (MR) \cite{milgram1994taxonomy}. A VR system blocks out the outside world and simulates a realistic environment where participants can move around and view, grab, and reshape objects in a virtual environment \cite{7786878}. As opposed to VR, AR provides a physical real-world environment that has been enhanced with 3D virtual objects \cite{8197491}, and MR is a hybrid reality where physical and digital objects coexist and interact in real-time \cite{zhang2018visualizing}. These XR technologies are redefining how people experience the world by blurring the line between the real and the virtual. It is expected that the global XR market will reach 300 billion U.S. dollars by 2024 \cite{xrmarket}.

\begin{figure}[htbp]
\centering
\includegraphics[width=3.2in]{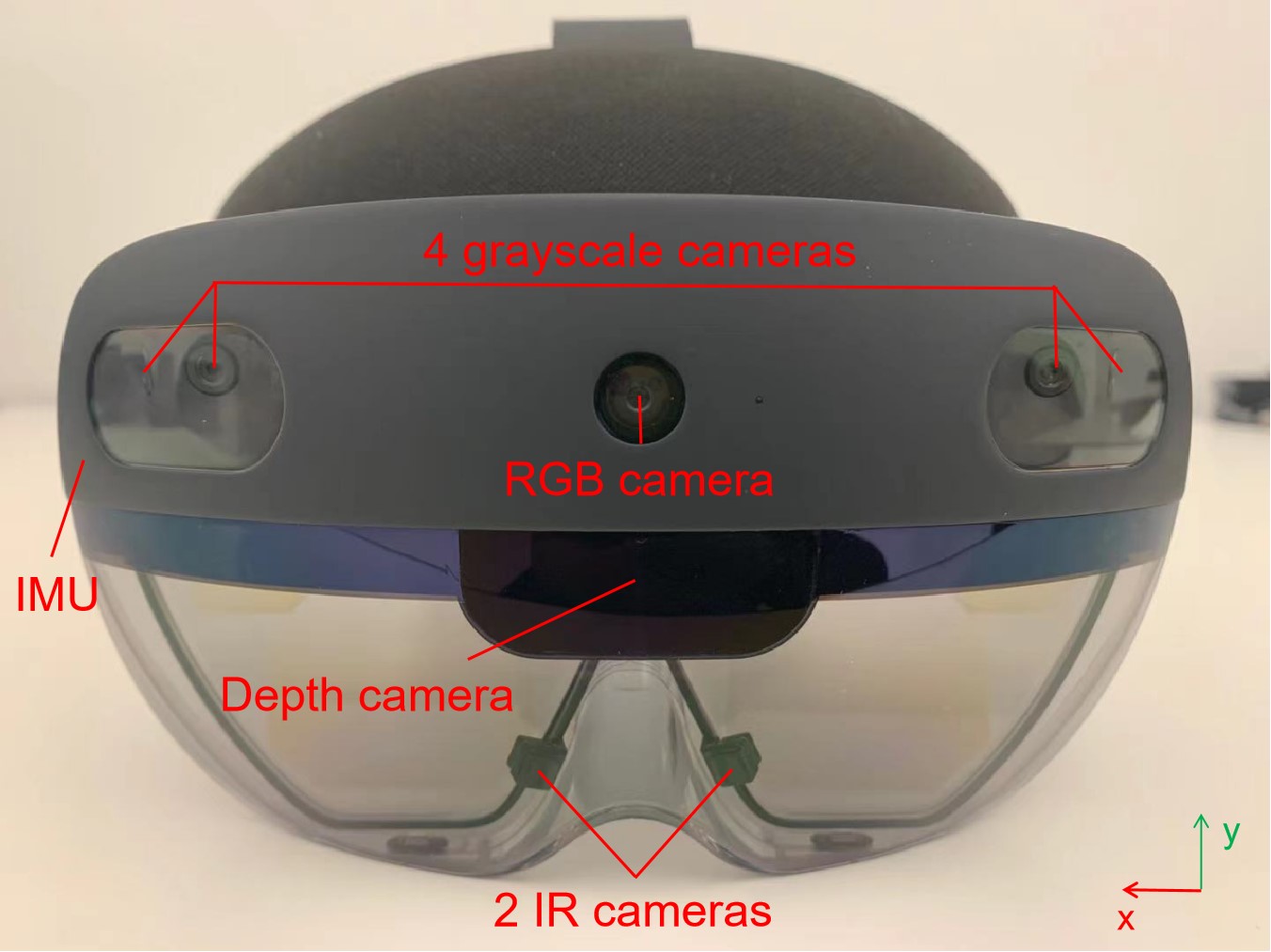}
\caption{HoloLens 2 is equipped with multiply sensors to support hand tracking, eye tracking, and voice command for human understanding. It is also capable of understanding environment by six degrees of freedom (6DoF) tracking and spatial mapping.}  
\label{hl2}
\end{figure}

\begin{table}[ht]
\centering
\caption{Technical specifications of cameras that the Microsoft HoloLens 2 are equipped with. These cameras introduce a large number of data for the HoloLens 2 to process to support these functions.}
\label{cameras}
\begin{adjustbox}{width=0.49\textwidth}
\begin{tabular}{|c | c | c | c | c | c|} 
\hline
& Cameras & FPS & Resolution & Format & Raw data \\ 
\hline
Head Tracking & 4 Grayscale & 30 & 640 x 480 & 8-bit & 294.9 Mbits\\
\hline
Hand Tracking & 1 Depth & 45 & 512 x 512 & 16-bit &  188.7 Mbits\\
\hline
Eye Tracking & 2 Infrared (IR) & - & - & - & -\\
\hline
Spatial Awareness & 1 Depth & 1-5 & 320 x 288 & 16-bit & 1.5-7.3 Mbits\\
\hline
\end{tabular}
\end{adjustbox}
{\raggedright (FPS stands for frames per second.)  \par}
\end{table}

\begin{figure*}[htbp]
\centering
\includegraphics[width=7.2in]{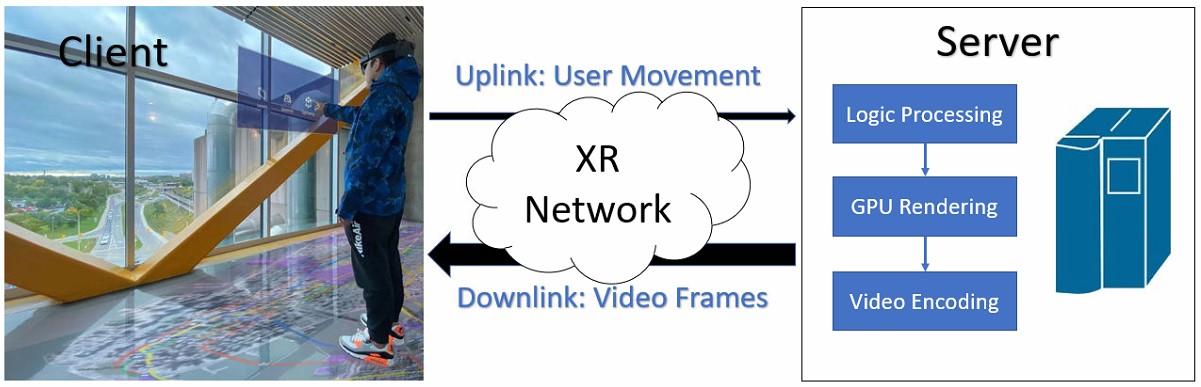}
\caption{The architecture of our proposed system consists of a user with the HoloLens 2, an XR network, and a rendering server. On the left, a user is wearing the HoloLens 2 to interact with the New York City data.}
\label{user}
\end{figure*}

Recently, we have seen an increase in the development of XR technologies, with an increasing number of companies producing XR devices.  A cutting-edge example of smart MR glasses is the Microsoft HoloLens 2. Compared to other MR devices, such as the HoloLens 1 \cite{liu2018technical}, the HoloLens 2 provides a more immersive experience with a larger Field of View (52°) and a higher resolution (2048×1080 per eye). As shown in Fig. \ref{hl2}, the HoloLens 2 employs eight cameras to facilitate human and environment understanding: four visible-light tracking cameras for real-time visual-inertial simultaneous localization and mapping (SLAM), a depth camera for articulated hand tracking as well as spatial awareness, two Infrared (IR) cameras for eye tracking, and an RGB color camera to create MR photos and videos for users. There are also Inertial Measurement Units (IMUs) including an accelerometer to determine the linear acceleration along the x, y, and z axis (right-hand rule), a gyroscope to determine rotations, and a magnetometer for absolute orientation estimation. These sensors lead to a large amount of data for the HoloLens 2 to transmit/process (Table \ref{cameras}). For example, for the four grayscale cameras used for head tracking, the raw data rate is 294.9 Mbits/s with a frame rate of 30 Hz and a resolution of 640x480 using 8 bits to encode the RGB data of each pixel. Therefore, computing power is in high demand to support all of these tracking functions.

Though the HoloLens 2 is equipped with a Holographic Processing Unit (HPU) to handle all computing tasks, the HPU may not be able to provide detailed models of specific scenarios where every detail matters, such as truck engines and city models. It is possible to solve these problems by remote rendering, which offloads complex rendering tasks to a rendering server with powerful computing capability. Users may experience a superior Quality of Experience (QoE) \cite{8509121} when using remotely rendered systems compared to those rendered by the HoloLens 2 itself. Fig. \ref{user} demonstrates the proposed XR system architecture in which a user interacts with the 3D digital data. The HoloLens 2 detects the user's movement and sends it to the remote server through a uplink (UL) stream of the XR network. The server performs logic processing and renders corresponding video frames based on the user's movement. The encoded video frames are then transmitted back to the user through a downlink (DL) stream of the XR network.

\begin{figure*}[htbp]
\centering
\includegraphics[width=7.2in ]{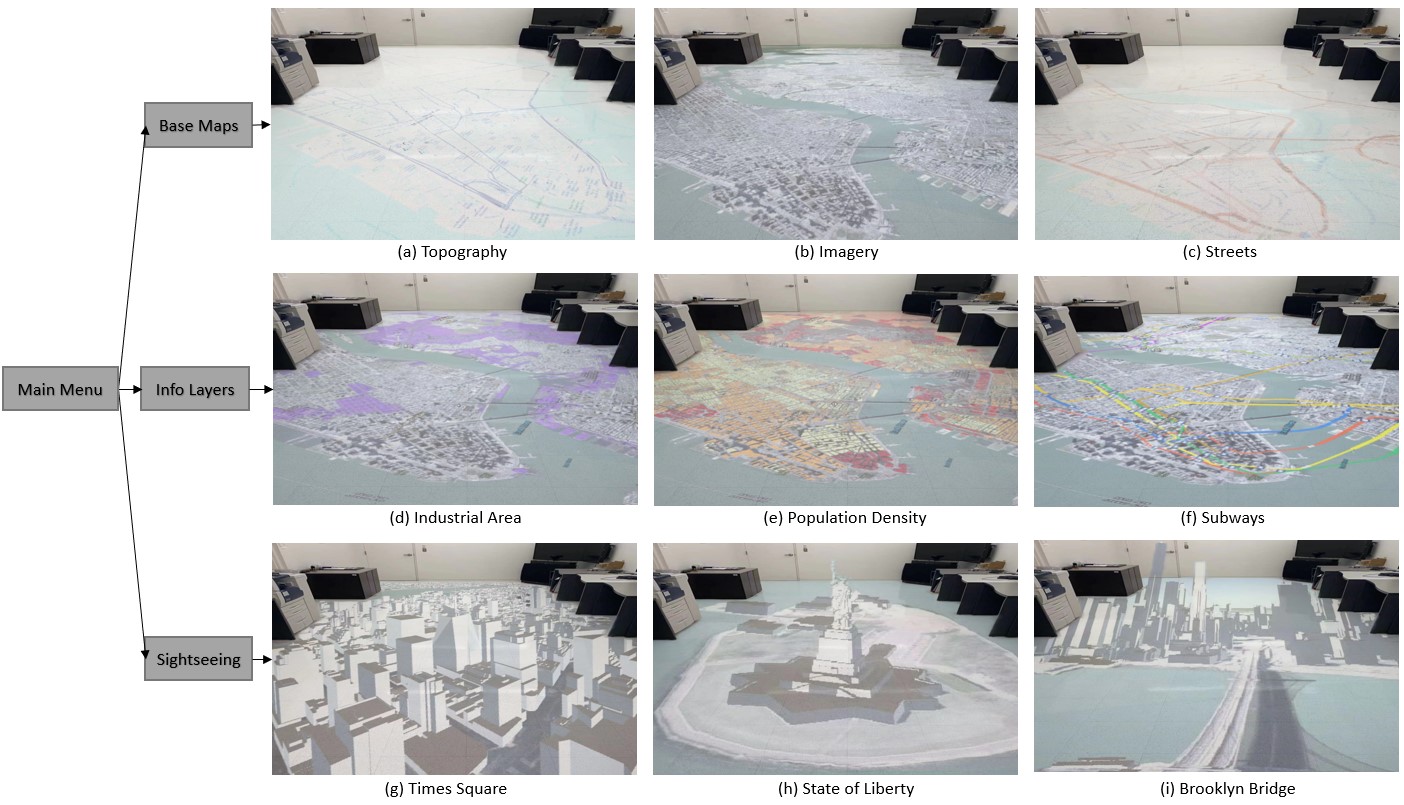}
\caption{There are three sections in the main menu: the ``Base Maps" section, the ``Info Layers" section, and the ``Sightseeing" section. We show three examples for each section: topography, imagery, streets for ``Base Maps", industrial area, population density, subways for ``Info Layers", and Times Square, Statue of Liberty, Brooklyn Bridge for ``Sightseeing".} 
\label{vis examples}
\end{figure*}

\section{Case Study: An XR System to Interact with Holographic Data of New York City}

\subsection{System Setup and Data Acquisition} 
Our rendering server is equipped with the Windows 10 operating system (OS), an Intel Core i9-11900F processor, 64 GB of random-access memory (RAM), and an NVIDIA GeForce RTX 3090 graphics card. For the HMD, the HoloLens 2 features the Qualcomm Snapdragon 850 Compute Platform, which comprises four gigabytes of LPDDR4x system memory. As a proof-of-concept, the Holographic Remoting Player is installed on the HoloLens 2 to connect the HoloLens 2 to the server through a USB-C cable. As an example made by Unity, we developed our system using multiple auxiliary packages. An important package is OpenXR, which is a royalty-free, open standard that provides high-performance access to XR platforms and devices. Nowadays, OpenXR has been supported by most of the industry-leading companies, such as Microsoft, Huawei, and Meta.

The New York City data was downloaded from ArcGIS Living Atlas of the World, the world's premier collection of geographic information \cite{atlas}. The site provides live feeds and other content that assists us in understanding current events and infrastructure in the form of information layers. These information layers contain geographic information such as maps and the distribution of buildings that can be imported directly into the system.

\subsection{Interaction Design}
We designed an interaction system where multiple interaction models are provided to allow users to naturally interact with holographic city data. Though there could be various effective and engaging interaction ways supported by the HoloLens 2 \cite{funk2017hololens}, we proposed three basic interaction models based on hand tracking, eye tracking, and voice commands in our system. With these natural interaction models, users can quickly learn how to manipulate objects within an XR environment without any difficulty.

\begin{figure}[ht] 
\centering
\subfigure[Pointing stage]{
\includegraphics[width=1.6in, height=1.6in]{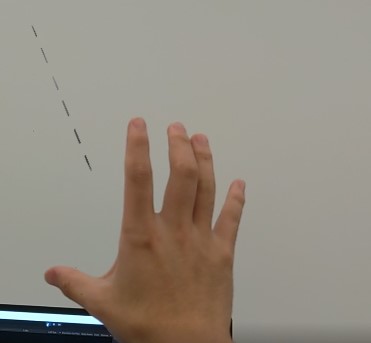}}
\subfigure[Committing stage]{
\includegraphics[width=1.6in, height=1.6in]{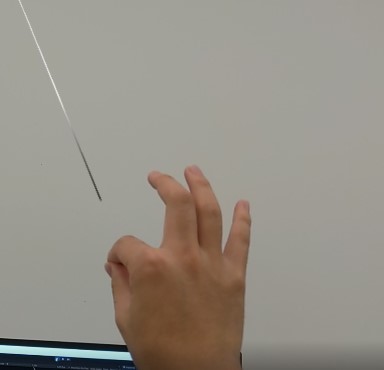}}
\caption{The far interaction of hand tracking is composed of two stages: the pointing stage to indicate the target object and the committing stage to manipulate the object.} 
\label{far}
\end{figure}

A key feature of the HoloLens 2 is hand tracking, which allows the device to identify the user's hands and fingers and track them in the air. With fully articulated hand tracking, users can touch, grasp, and move holographic objects naturally. In our system, hand tracking supports two types of interactions: the near interaction and the far interaction. The near interaction is a method of controlling objects that are physically close to the user, allowing their hands to directly manipulate them. As an example, buttons are activated by simply pressing them. In addition, we provide a way to manipulate objects out of reach using the ``point and commit" interaction, which is a unique interaction method in the XR world. It involves the ray shooting out of the user's palm and includes two stages: the pointing stage and the committing stage. In the former stage, users should reach out their hands with a dashed ray (Fig. \ref{far} (a)) while the end of the ray indicates the target object. In the latter stage, which can be triggered by the thumb and index finger (Fig. \ref{far} (b)), the ray becomes a solid line, allowing users to interact with 3D objects from a distance. 

The HoloLens 2 is able to detect a user's gaze by using sensors observing at the eyes. This eye gaze indicates a signal for the user's focus and intent. Moreover, it is important for users to receive appropriate feedback and indication when interacting with the city data in our system. As an example, when a user gazes at a famous building in New York City in our system, the name of the building will appear as feedback. The other hands-free modality in our system to interact with holograms is using voice commands. Voice commands are usually used in conjunction with eye gaze as a method of targeting. For example, when the user is looking at a button, the button can be activated by saying ``Open". We provide the user feedback with the word ``Open" showing up in front of the user for 2 seconds. We also allow users to directly say the names of sightseeing in the city model, such as ``Times Square" to show the views of Times Square.


\subsection{Visualization Results}
As shown in Fig. \ref{vis examples}, we describe the architecture of our system with several visualization examples. The main menu consists of three sections: the ``Base Maps" section, the ``Info Layers" section, and the ``Sightseeing" section. In our scenario, a base map is used to provide a background of the geographic context of our city model. We provide three types of base maps including topography base maps (Fig. \ref{vis examples} (a)), imagery base maps (Fig. \ref{vis examples} (b)), and streets base maps (Fig. \ref{vis examples} (c)). The topographic base map provides a detailed and accurate representation of the elements on the surface of the earth.  An imagery base map includes satellite imagery for the world, along with high-resolution aerial imagery for many areas, whereas a streets base map shows road and transit network details that are legible and accurate.  The ``Info Layers" section provides geographical information that can be displayed as tiles on the base map. We divide all the info layers into three aspects: Environment, Infrastructure, and People. Users have a variety of options to choose from in each aspect. For example, we provide the industrial area (Fig. \ref{vis examples} (d)), population density (Fig. \ref{vis examples} (e)), and poverty distribution in New York City in the ``People” related layers and subways (Fig. \ref{vis examples} (f)), buildings, and transit frequency situation in the ``Infrastructure” related layers. In our ``Sightseeing” section, a few of well-known attractions are displayed as options such as Times Square (Fig. \ref{vis examples} (g)), Statue of Liberty (Fig. \ref{vis examples} (h)), Brooklyn Bridge (Fig. \ref{vis examples} (i)), and Central Park. When the base map with info layers or one of the attractions is selected, users will be able to see the detailed views.

\subsection{Quality of Experience (QoE) Metric Design}
Modeling the QoE of XR systems has been extensively researched \cite{8509121}. However, these QoE models are primarily designed for evaluating performances where frames are locally rendered by XR devices. In remote rendering XR systems, users are highly sensitive to latency, due to the delay introduced by video encoding, network delay, and video decoding, which are necessary steps for real-time rendered video streaming from the server. Unfortunately, few research studies have been conducted on QoE metrics for remotely rendered XR systems. In this paper, we propose a time-window based QoE model for this purpose which is defined as:
\begin{equation}
QoE=\sum_{n=1}^{N}q(F_n) + \sum_{n=1}^{N}p(R_n) - u \sum_{n=1}^{N}g(L_n)
\end{equation}
where $q(F_n)$ represents the level of a user's satisfaction with the smoothness of the rendered video, where $F_n$ donates the average frame rate at time window $n$. The level of satisfaction with video quality is indicated by $p(R_n)$, where $R_n$ represents the average resolution at time window $n$. $g(L_n)$ is used to penalize the turnaround latency between sending pose data from XR devices to the remote server and displaying the video frame for that pose data on the display, where $L_n$ donates the average latency at time window $n$. Because the marginal improvement in perceived quality decreases with higher frame rates and resolutions \cite{9110784}, we used two logarithmic functions to represent $q(F_n)$ and $p(R_n)$, where $q(F_n)=log(F_n/F_{min})$ and $p(R_n)=log(R_n/R_{min})$. $F_{min}$ and $R_{min}$ are the minimum values of the frame rates and the resolutions. Conversely, a user's satisfaction significantly decreases more as the total latency increases. Therefore, we used an exponential function to denote $g(L_n)$, i.e. $g(L_n)=e^{L_n/L_{min}}$ and $u$ is the latency penalty factor.

\subsection{Remote Rendering vs. Local Rendering}
We evaluated our proposed remote rendering enabled XR system by comparing its performances with local rendering by the HoloLens 2 itself. For both cases (remote rendering and local rendering), the system is required to achieve 60 FPS with a resolution of 2048x1080 and a latency of approximately 60 milliseconds (ms) to provide a satisfactory experience \cite{liu2018cutting}. In our experiments, we compared the performances of the systems in both cases using the above parameters as the target frame rate and the target resolution. We also evaluated the case when the target resolution is 1024x540 to further investigate their performances in high-quality resolutions and low-quality resolutions, respectively.

\begin{figure}[htbp]
\centering
\subfigure[Frame rate]{
\includegraphics[width=0.5\textwidth]{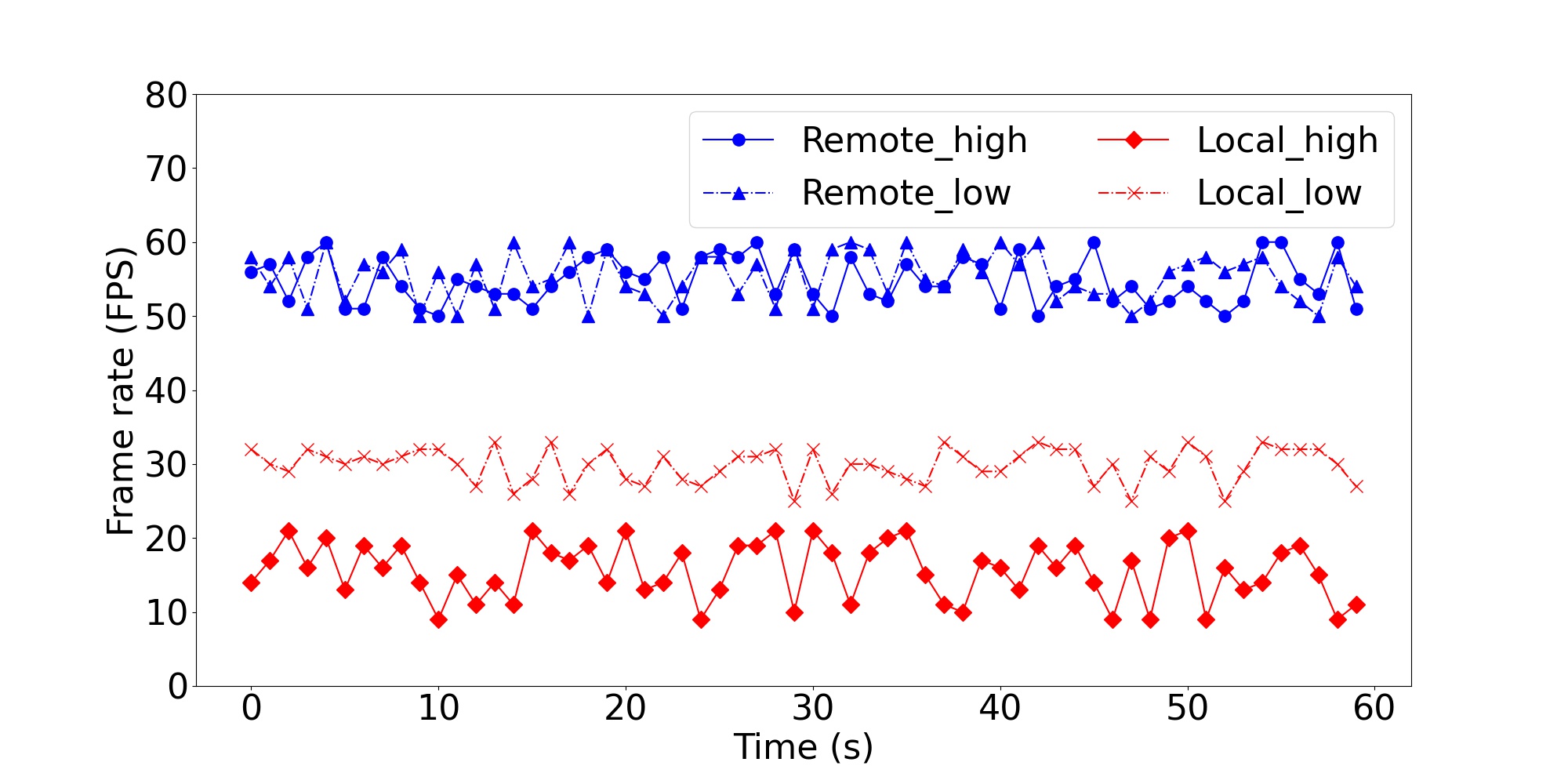}}
\subfigure[Latency]{
\includegraphics[width=0.5\textwidth]{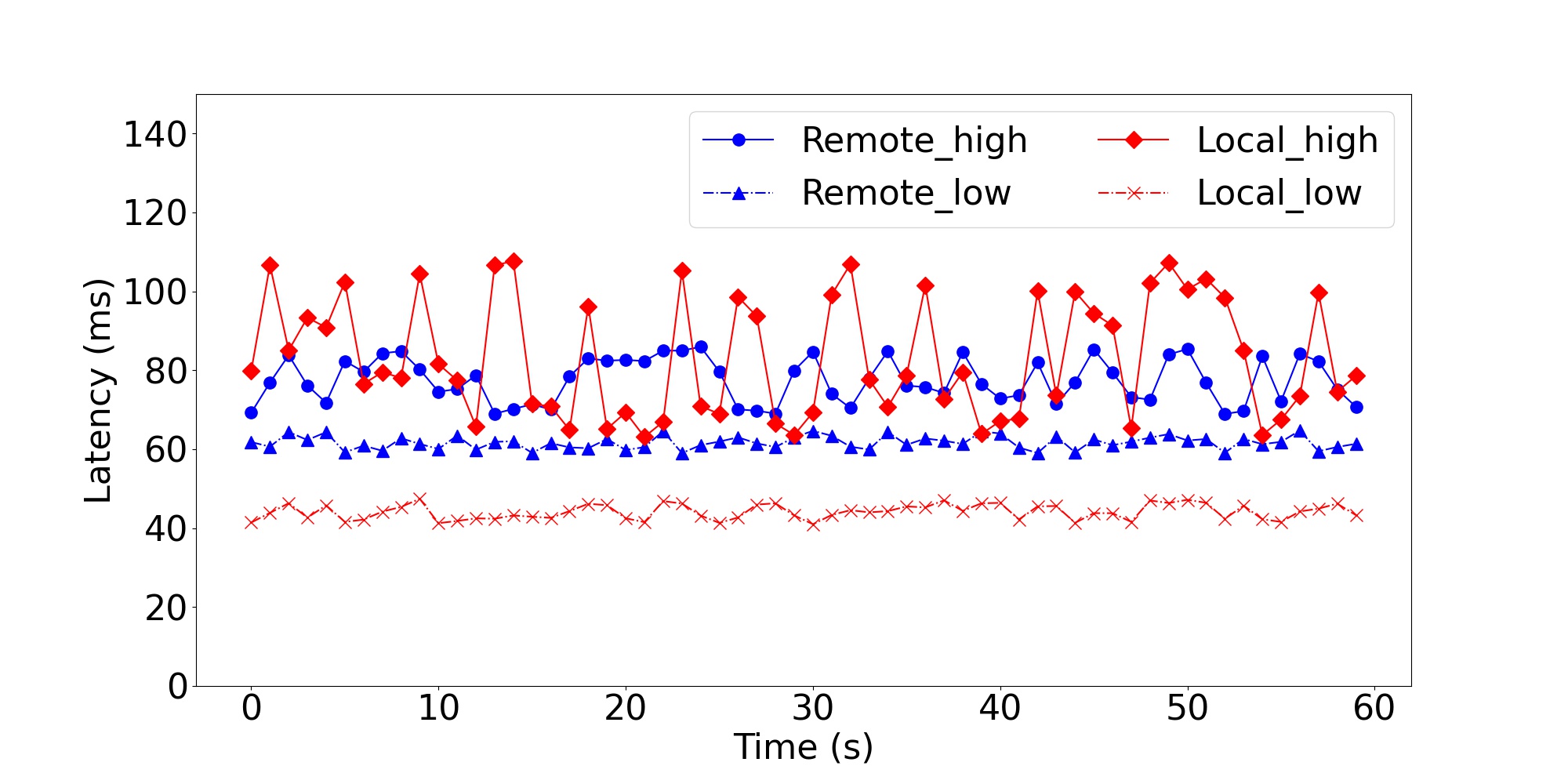}}
\subfigure[QoE]{
\includegraphics[width=0.5\textwidth]{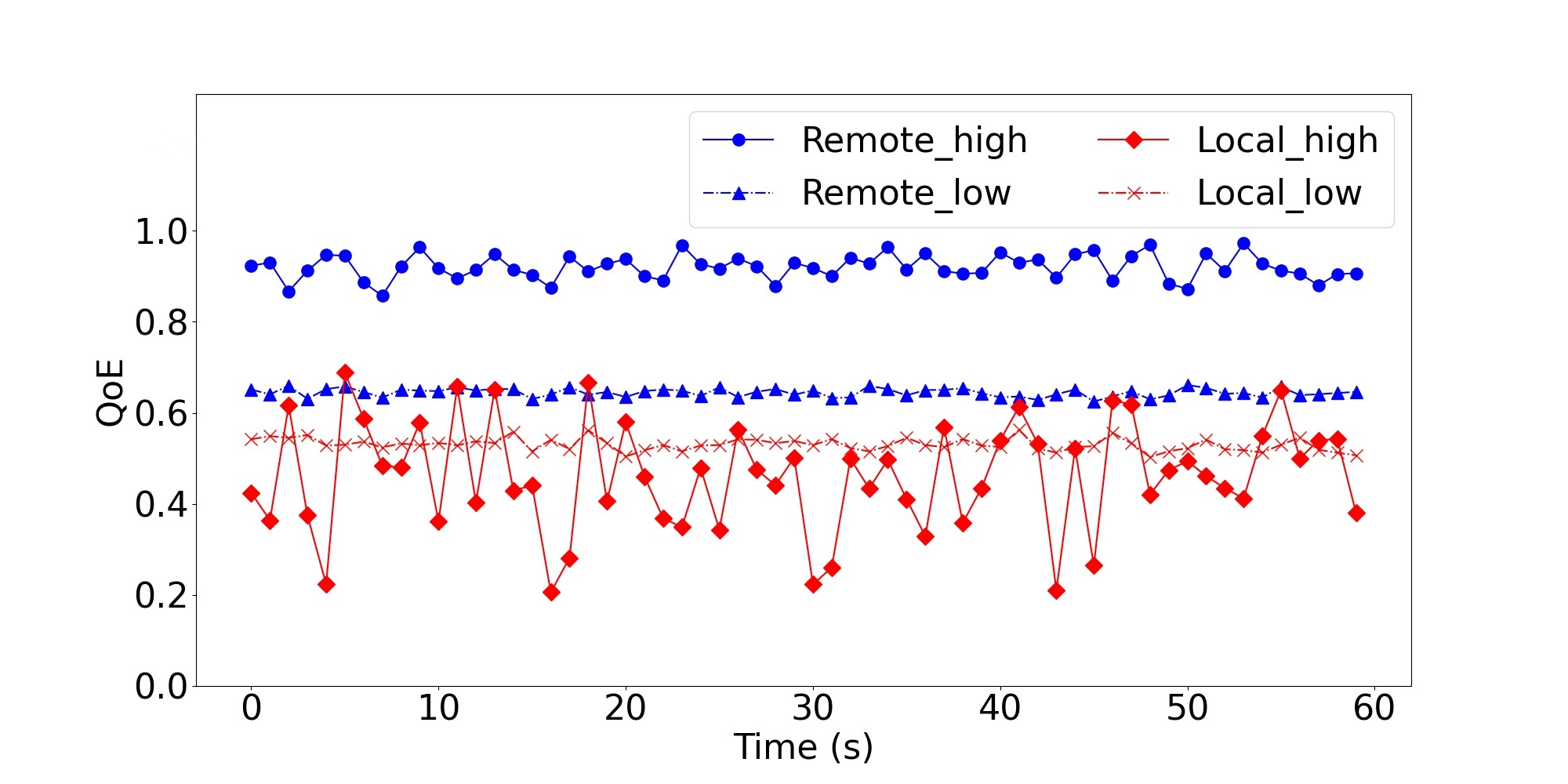}}

\caption{The comparison of our proposed remotely rendered system with the locally rendered system in terms of (a) frame rate, (b) latency, and (c) QoE for high-quality resolution (2048x1080) and low-quality resolution (1024x540). The QoE calculation equation is presented in Eq.1.} 
\label{vs}
\end{figure}

Four scenarios are tested and compared in terms of frame rate, latency, and QoE (as shown in Fig. \ref{vs}), including the remotely rendered high-quality videos (denoted as Remote\_high), the remotely rendered low-quality videos (Remote\_low), the locally rendered high-quality videos (Local\_high) and the locally rendered low-quality videos (Local\_low). We observe that the frame rates are far below the required level (60 FPS) when rendering the New York City model locally. For the Local\_high videos, the frame rates are scattered from 9 to 21 FPS, while for the Local\_low videos, the frame rate is a little better, but still quite low (around 25-33 FPS). In contrast, the frame rates of remote rendering are stable and maintain around 55-60 for both the Remote\_high videos and the Remote\_low videos. The reason might be that a considerable amount of time is spent on rendering frames by the GPU on the HoloLens 2 compared to remote rendering. The smoothness of the rendered videos is remarkably improved on the powerful rendering server. Regarding latency, the Local\_low videos perform the best (around 40 ms). However, for the Local\_high videos, the latency can be as high as 110 ms which may lead to sickness. Although remotely rendered systems introduce extra latency (such as encoding latency, transmission latency, and decoding delay), the total latency is much lower than the Local\_high videos: 59-65 ms for the Remote\_low videos and 68-86 ms for the Remote\_high videos. In regard to QoE defined in Eq. 1, Remote\_high outperforms the other three scenarios at each time window, with significant improvement in the average QoE compared to Local\_high (0.92 vs 0.46). Remote\_low also improves the average QoE by 21\% compared to Local\_low (0.64 vs 0.53). Due to the limited computing resources on the HoloLens 2, the QoE scores for the Local\_high videos are quite low and unstable.

\section{Network Traffic Analysis of the Proposed XR System}
\subsection{XR Network}
XR networks \cite{9539162} is defined as a network that carries XR contents. Specifically, in our system, the XR network connects the HoloLens 2 and the rendering server. To analyze the characteristics of the XR network in our system, we ran Wireshark, a packet sniffer, on the rendering server to capture the network traffic. The traffic analysis was performed at the default frame rate of the HoloLens 2 (60 FPS), with the default resolution of the HoloLens 2 (2048×1080) High-Efficiency Video Coding (HEVC) as the compression standard and no limitation on the data rate. The captured network packets were stored in our local databases. We decoded the packets by Scapy, a Python library for manipulating network packets, to acquire key information such as protocols in use and packet sizes. We found in our system that User Datagram Protocol (UDP) sockets over IPv4 are used instead of Transmission Control Protocol (TCP). The reason could be that UDP is faster, simpler, and more efficient than TCP for XR systems where latency has a higher priority than a small amount of data loss.

\subsection{XR Network Traffic Modeling}
To design an effective and high-performance XR network, it is necessary to accurately characterize and model the traffic of XR networks. Modeling XR network traffic can also be highly useful for simulation studies, and generation of synthetic XR traffic traces for testing \cite{9539162}. Among the various characteristics of XR networks, the following two are of particular interest: the distributions of video frame data and video frame size prediction.

\subsubsection{Distributions of video frame data}
\begin{figure}[htbp]
\centering
\includegraphics[width=0.5\textwidth]{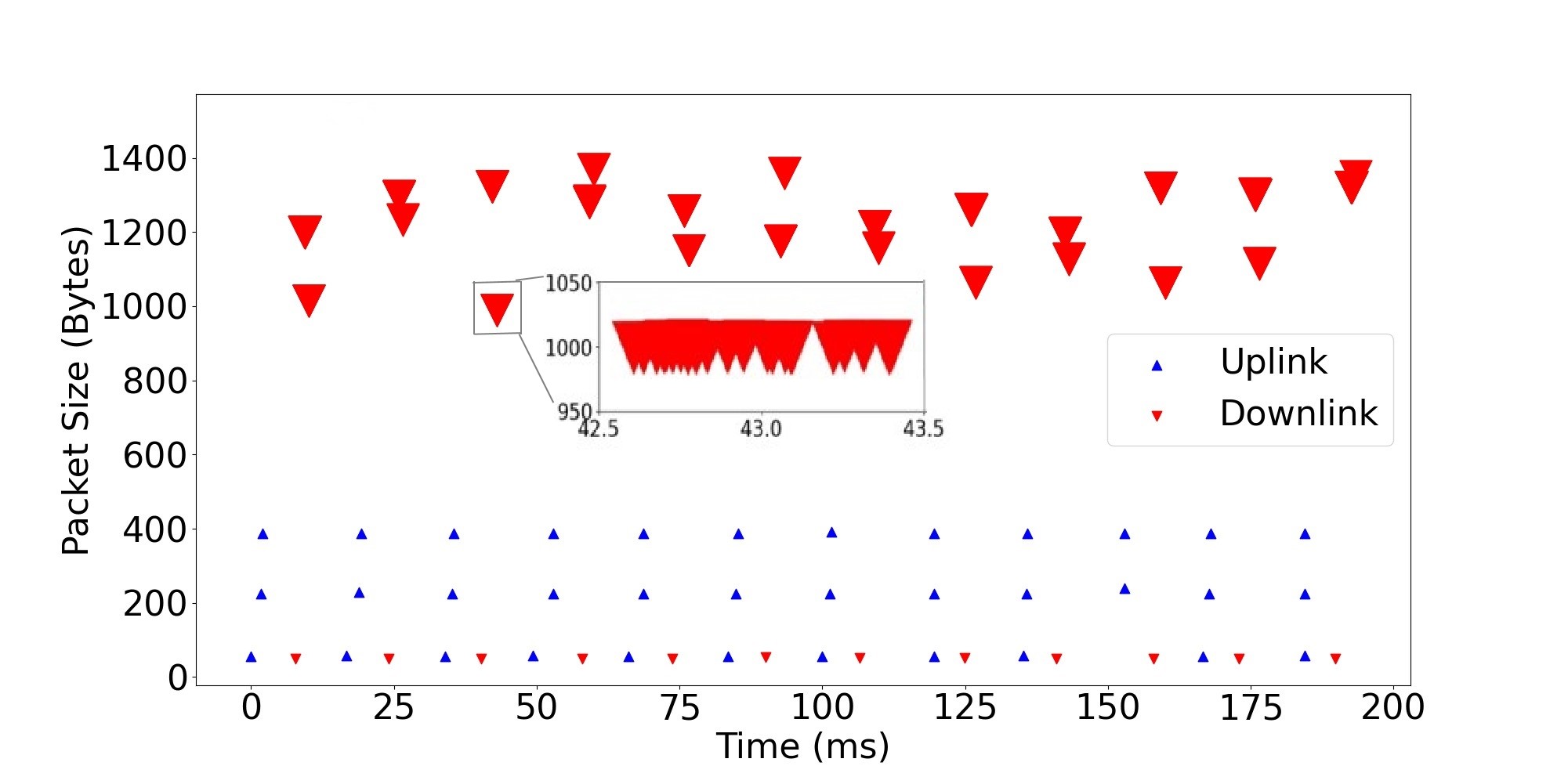}
\caption{Part of the captured packets from the UL and the DL stream in the XR network. A video frame is transmitted by two packet bursts for two eyes. Each burst contains 8-55 long packets with more than 1000 Bytes.} 
\label{All stream}
\end{figure}

We illustrate a portion of the bidirectional network traffic generated by our system as shown in Fig. \ref{All stream}. In the UL stream, we can see that two or three packets are sent almost simultaneously from the HoloLens 2 every 17 ms, which is similar to the time interval between two frames. Because the frame rate is 60 FPS, the expected time for a frame is 16.7 ms. We assume that these packets are mainly used as user data and synchronization information for our system to render frames. For the DL stream which contains rendered frames and synchronization information, there are two types of packets with different sizes. Long packets with more than 1000 Bytes are sent in bursts of multiple packets and every 17 ms two adjacent bursts are sent. We believe this is because the rendering server sends back the rendered frames for two eyes by two close bursts every 17 ms. Short packets containing synchronization messages from our system are sent separately at the same interval. Therefore, the frequency of data transmission between the HoloLens 2 and the remote system depends on the number of frames displayed on the HoloLens 2 per second. 

\begin{figure}[ht] 
\centering
\subfigure[Distribution of frame size]{
\includegraphics[width=0.23\textwidth, height=1in]{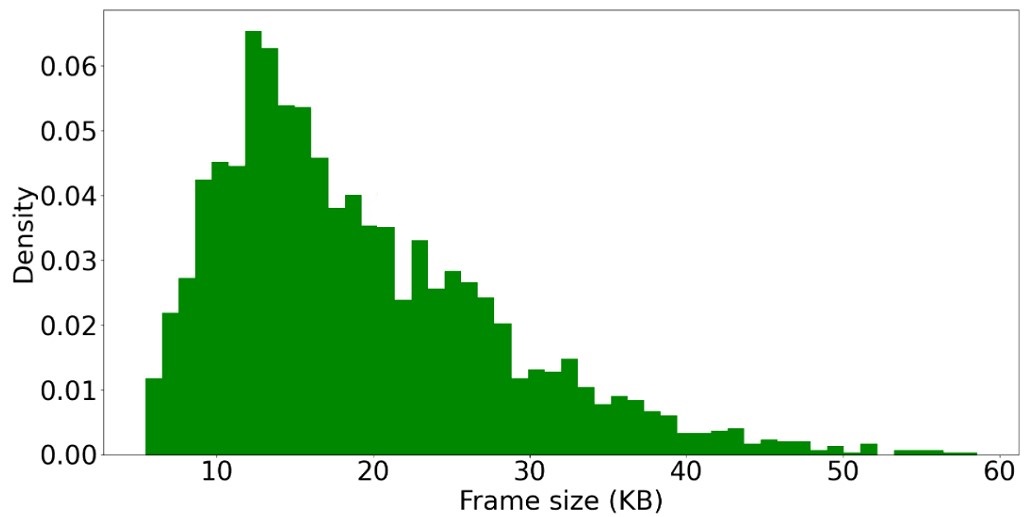}}
\subfigure[Q-Q plot of frame size]{
\includegraphics[width=0.23\textwidth, height=1in]{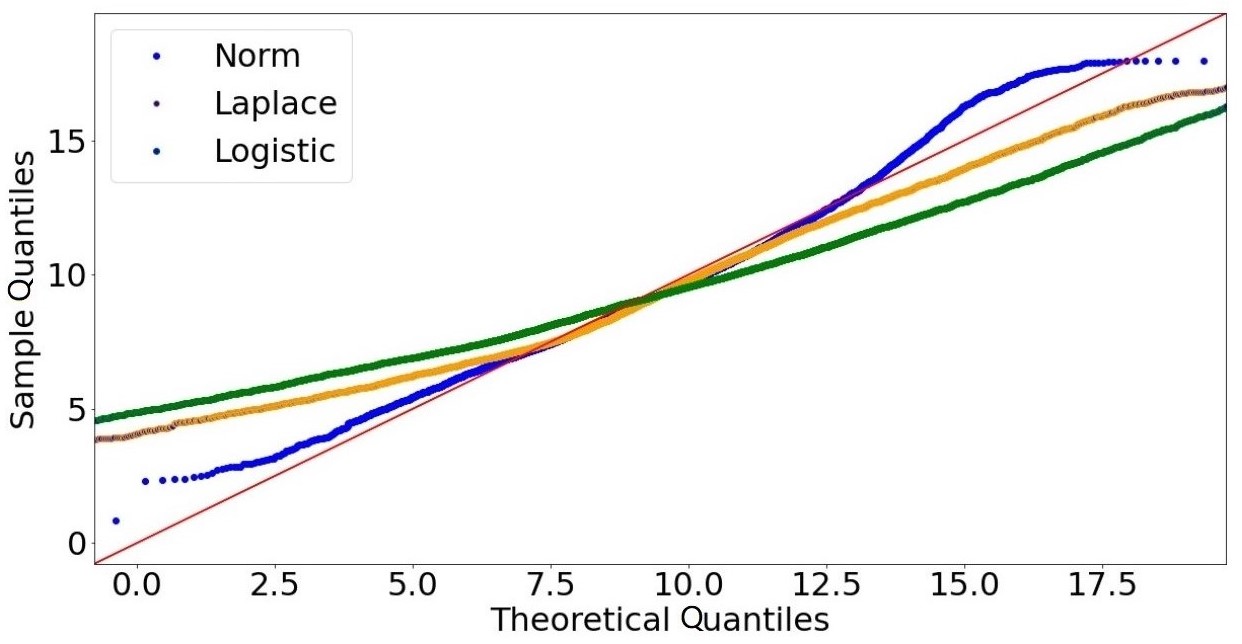}}
\subfigure[Distribution of frame interval]{
\includegraphics[width=0.23\textwidth, height=1in]{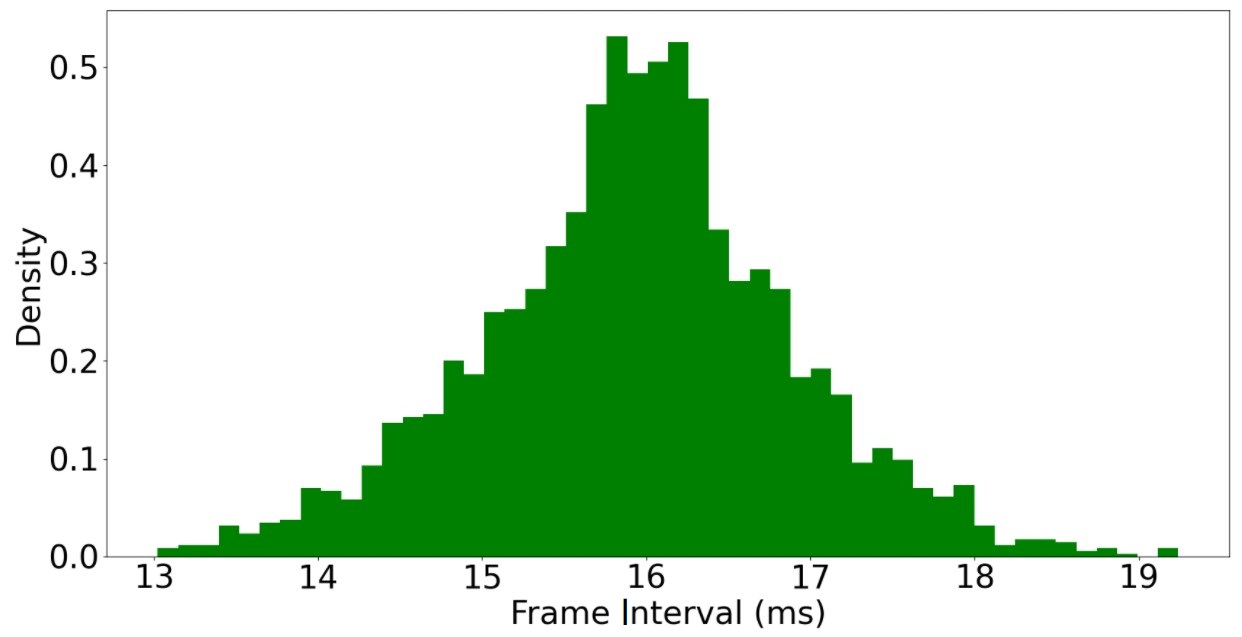}}
\subfigure[Q-Q plot of frame interval]{
\includegraphics[width=0.23\textwidth, height=1in]{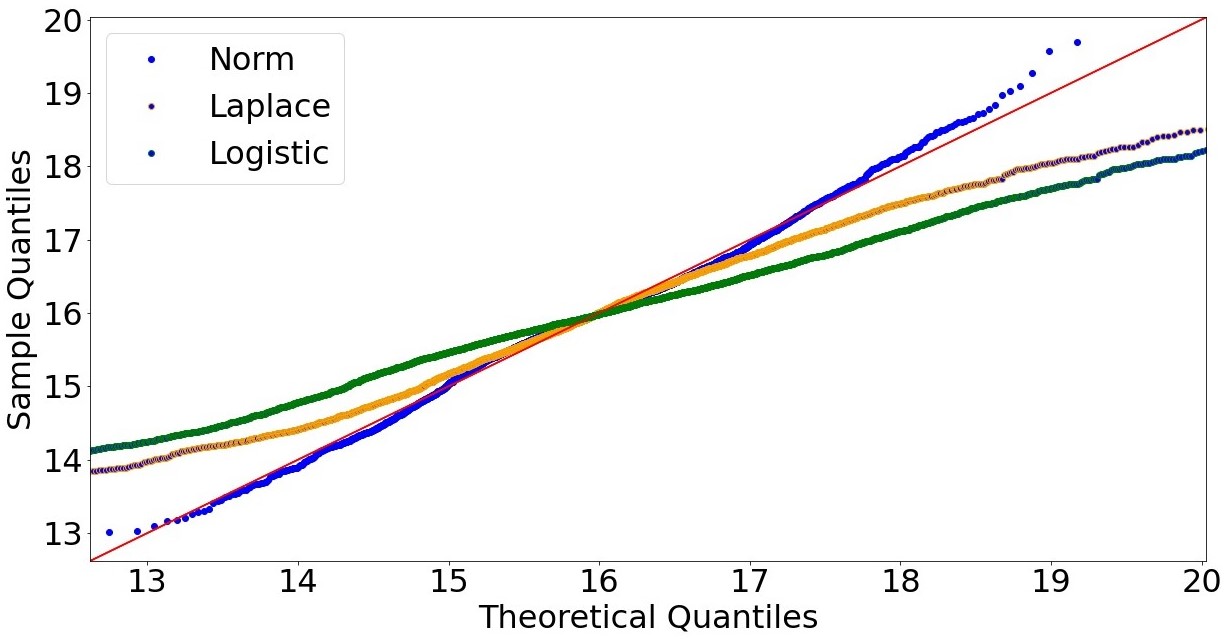}}
\subfigure[Distribution of eye interval]{
\includegraphics[width=0.23\textwidth, height=1in]{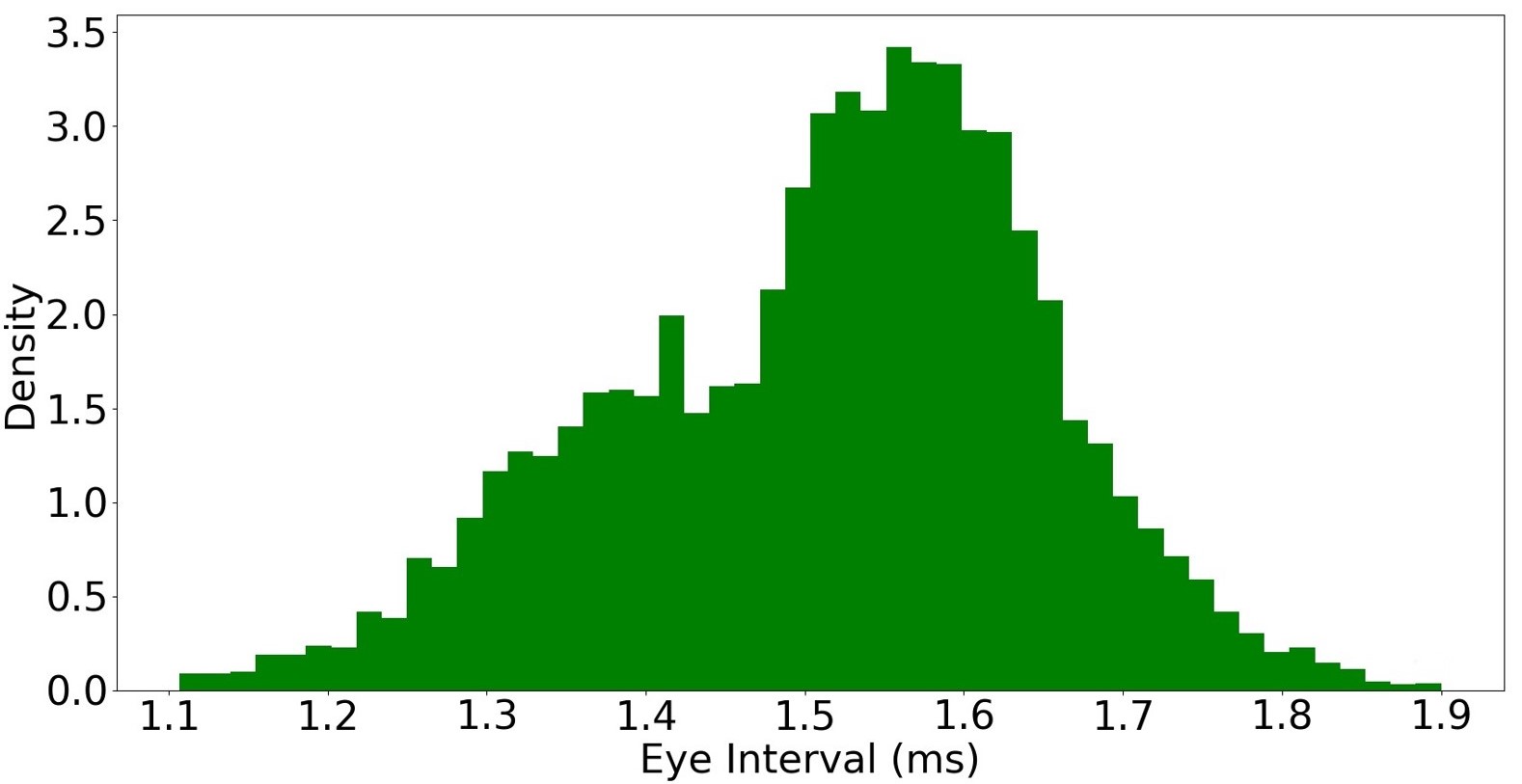}}
\subfigure[Q-Q plot of eye interval]{
\includegraphics[width=0.23\textwidth, height=1in]{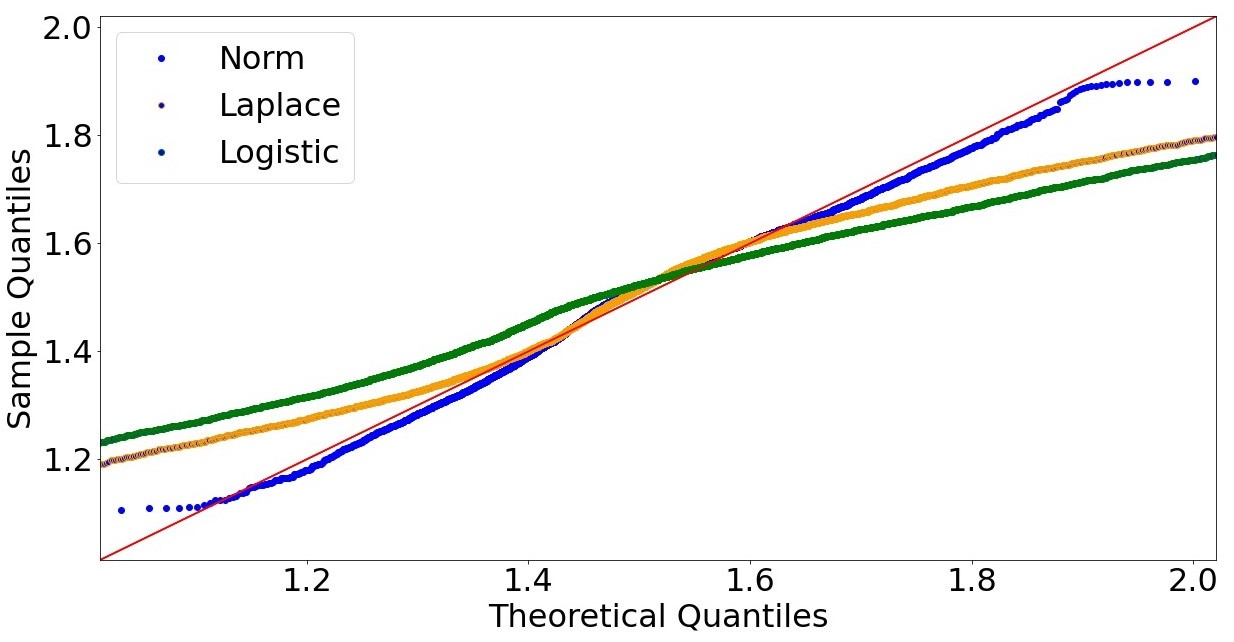}}
\caption{Distributions of the video frame size, the frame interval, and the eye interval. We employed the Q-Q plots where we compared our sample data to the normal distribution, the Laplace distribution, and the logistic distribution. } 
\label{models}
\end{figure}

We focused on the distributions of the video frame size, the frame arrival interval, and the arrival interval between two bursts for two eyes (the eye interval). As shown in Fig. \ref{models} (a), we demonstrate the distribution of the video frame size using histograms. We constructed a quantile-quantile (Q-Q) plot where the sample points fall approximately on the line y = x if the two distributions being compared are similar \cite{9539162}. To see how the sample data fits different distributions, we compared the sample data to the normal distribution ($Normal(x)={e^{-\frac{x^2}{2}}}/{\sqrt{2\pi}}$), the Laplace distribution ($Laplace(x)={e^{-|x|}}/{2}$), and the logistic distribution ($Logistic(x)={e^{-x}}/{(1+{e^{-x}})^2}$), with Probability Density Functions (PDFs) in their standardized forms.

As shown in Fig. \ref{models} (b), the sample data points for the normal distribution almost lie on the line y = x, indicating that the frame size can be considered to be subject to the normal distribution. Moreover, the sample data fits better on a normal distribution compared to the other two distributions as the line for the normal distribution lies closer to the line y = x. We can get the similar conclusions for the distributions of the frame interval (shown in Fig. \ref{models} (c), (d)) and the eye interval (shown in Fig. \ref{models} (e), (f)). Therefore, we summarize that the frame size, the frame interval, and the eye interval of our system are all subject to the normal distribution.
\subsubsection{Video frame size prediction}
\begin{figure}[htbp]
\centering
\includegraphics[width=0.5\textwidth]{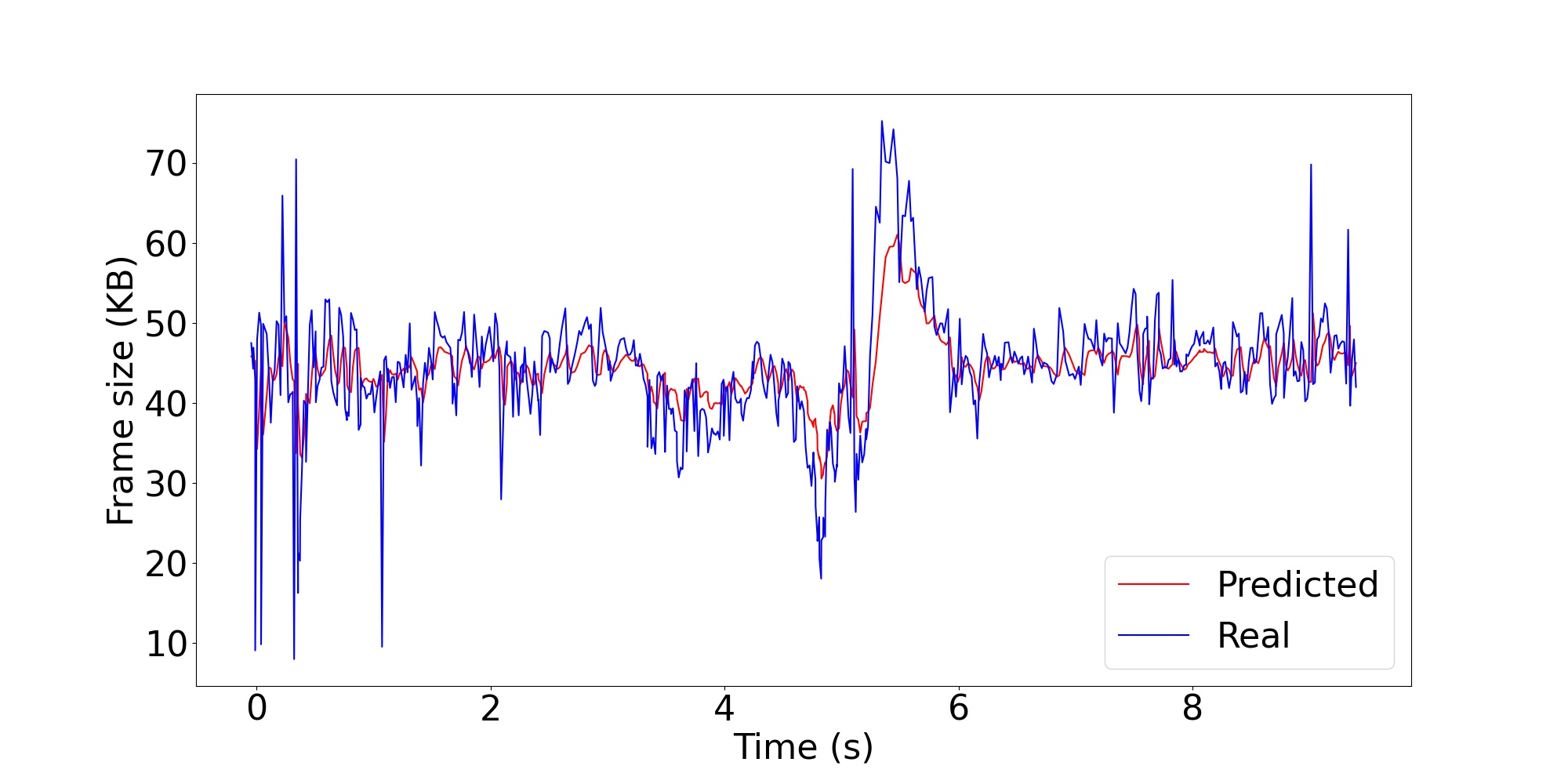}
\caption{The predicted frame sizes by our ARMA-based model compared with the ground truth of the frame sizes on the test data set.} 
\label{pred}
\end{figure}

Predicting the oncoming video frame sizes is challenging because XR network traces exhibit both Long Range Dependent (LRD) and short-range dependent (SRD) properties. Several models have been proposed for traditional HTTP-based video traffic in the literature \cite{6422289}. However, XR network traffic typically uses the Real-time Transport Protocol (RTP) instead of the Hypertext Transfer Protocol (HTTP) as its protocol and displays several different characteristics such as elephant DL stream and mice UL stream compared with traditional video traffic. To the best of our knowledge, there is little work on XR traffic prediction. We denoted the video frame size by a time series of data $F_t$ and proposed an Autoregressive Moving Average (ARMA) model for predicting future values in the series. The model is referred as the ARMA(p, q) model where p is the order of the Autoregressive (AR) part and q is the order of the Moving Average (MA) part, defined as:
\begin{equation}
F_t=c + \epsilon_t + \sum_{i=1}^{p}\phi_iF_{t-i} + \sum_{i=1}^{q}\theta_i\epsilon_{t-i}
\end{equation}
where $\phi _{i},...,\phi _{p}$ are parameters of the AR model, $\theta _{i},...,\theta_{q}$ are parameters of the MA model, $\epsilon_{t},...,\epsilon_{t-q}$ are white noise error terms and $c$ is a constant. Because a stationary time series whose properties do not depend on time is easier to predict by an ARMA model, we first checked the stationarity of the video frame size series using the augmented Dickey-Fuller (ADF) test. The results indicated that the series is stable with a 95 percent confidence interval. We then drew an autocorrelation factor (ACF) plot and a partial autocorrelation factor (PACF) plot to determine the parameters of the ARMA model (based on our analysis, p is 5 and q is 4). In our experiment, 70 percent of the data was used to train the model and the remaining 30 percent was used to evaluate its accuracy. We observe that the differences between the predicted values and the real values in the testing are not big (as shown in Fig. \ref{pred}) which verifies that the proposed model captures the autocorrelation structure of the frame sizes even when dramatic fluctuations in the frame size occur (between 4-6 s).

\subsection{Discussion and Future Work}
There are still other challenges that we have not solved yet. It is noted that latency plays an important role in XR systems. The round-trip latency should be less than 20 ms for the Motion-To-Photon latency (MTP) to become imperceptible \cite{ju2017ultra}. If the latency is too large, users may experience symptoms of motion sickness \cite{chang2020virtual}. However, using a remote rendering server introduces extra latency for encoding frames, transmitting encoded frames by networks, and decoding frames. Another important challenge of remote rendering on the HoloLens 2 is the throughput of networks for transmitting high-quality video frames. For the HoloLens 2 with a frame rate of 60 Hz and a resolution of 2048×1080, if 24 bits are used for a single pixel, the raw data rate would be 2.96 Gbps \cite{liu2018cutting}. Therefore, the future work is to figure out a way to lower the latency and increase the capacity for our XR network.

\section{Conclusion}
In this article, we demonstrated a remote rendering system for interacting high-quality holographic city data with the HoloLens 2 in real-time. We evaluated the proposed system by comparing its performances with local rendering by the HoloLens 2 itself in terms of frame rate, latency, and QoE. The results indicated that the proposed system can significantly improve QoE by at least 21\% under different resolutions. In addition, the traffic characteristics of the proposed XR network, such as the frame size, the frame interval, and the eye interval are analyzed and modeled which would benefit the design of any remote rendering systems based on the OpenXR standard.


%





\ifCLASSOPTIONcaptionsoff
  \newpage
\fi





\bibliographystyle{IEEEtran}
\bibliography{Bibliography}
%


\begin{IEEEbiography}[{\includegraphics[width=1in,height=1.25in,clip,keepaspectratio]{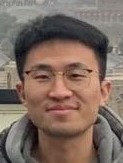}}]{Zijian Long}
(zlong038@uottawa.ca) received the B.Sc. degree in Software Engineering from Beijing Institute of Technology, China, in 2016 and the M.Sc. degree in Electrical and Computer Engineering from the University of Ottawa, Canada, in 2020. He is currently a PhD candidate in the School of Electrical Engineering and Computer Science, University of Ottawa. His research interests include XR network and artificial intelligence.
\end{IEEEbiography}

\begin{IEEEbiography}[{\includegraphics[width=1in,height=1.25in,clip,keepaspectratio]{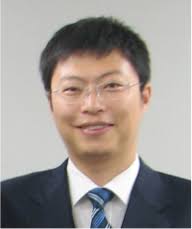}}]{Haiwei Dong}
(haiwei.dong@ieee.org) received the Ph.D. degree in computer science and systems engineering from Kobe University, Kobe, Japan in 2010 and the M.Eng. degree in control theory and control engineering from Shanghai Jiao Tong University, Shanghai, China, in 2008. He was a Principal Engineer with Artificial Intelligence Competency Center in Huawei Technologies Canada, Toronto, ON, Canada, a Research Scientist with the University of Ottawa, Ottawa, ON, Canada, a Postdoctoral Fellow with New York University, New York City, NY, USA, a Research Associate with the University of Toronto, Toronto, ON, Canada, and a Research Fellow (PD) with the Japan Society for the Promotion of Science, Tokyo, Japan. He is currently a Principal Researcher with Waterloo Research Center, Huawei Technologies Canada, Waterloo, ON, Canada, and a registered Professional Engineer in Ontario. His research interests include artificial intelligence, multimedia communication, multimedia computing, and robotics. He also serves as a Column Editor of IEEE Multimedia Magazine and an Associate Editor of ACM Transactions on Multimedia Computing, Communications, and Applications.
\end{IEEEbiography}


\begin{IEEEbiography}[{\includegraphics[width=1in,height=1.25in,clip,keepaspectratio]{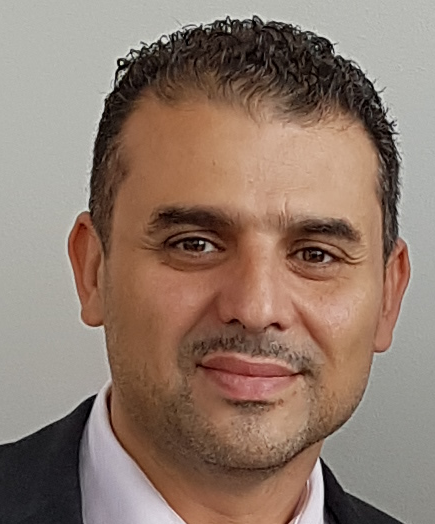}}]{Abdulmotaleb El Saddik}
(elsaddik@uottawa.ca) is currently a Distinguished Professor with the School of Electrical Engineering and Computer Science, University of Ottawa and a Professor with Mohamed bin Zayed University of Artificial Intelligence. He has supervised more than 120 researchers. He has coauthored ten books and more than 550 publications and chaired more than 50 conferences and workshops. His research interests include the establishment of digital twins to facilitate the well-being of citizens using AI, the IoT, AR/VR, and 5G to allow people to interact in real time with one another as well as with their smart digital representations. He received research grants and contracts totaling more than \$20 M. He is a Fellow of Royal Society of Canada, a Fellow of IEEE, an ACM Distinguished Scientist and a Fellow of the Engineering Institute of Canada and the Canadian Academy of Engineers. He received several international awards, such as the IEEE I\&M Technical Achievement Award, the IEEE Canada C.C. Gotlieb (Computer) Medal, and the A.G.L. McNaughton Gold Medal for important contributions to the field of computer engineering and science.
\end{IEEEbiography}






\end{document}